\documentclass[preprint,preprintnumbers,amsmath,amssymb]{revtex4}
\usepackage{amsfonts}    
\usepackage{amssymb}
\usepackage{latexsym}
\usepackage{graphicx}
\usepackage[usenames,dvipsnames]{color}

\newcommand{\pa}{\partial}

\newcommand{\ta}{\tau}

\newcommand{\ka}{\kappa}
\newcommand{\De}{\Delta}

\newcommand{\half}{\frac{1}{2}}
\newcommand{\rar}{\rightarrow}

\begin{document}

\title{The $BC_{1}$ quantum Elliptic model: algebraic forms, hidden algebra $sl(2)$, polynomial eigenfunctions}

\author{A.V.~Turbiner}
\email{turbiner@nucleares.unam.mx}
 \affiliation{Instituto de Ciencias Nucleares, Universidad Nacional
 Aut\'onoma de M\'exico, Apartado Postal 70-543, 04510 M\'exico,
 D.F., Mexico}

\begin{abstract}
The potential of the $BC_1$ quantum elliptic model is a superposition of two Weierstrass functions with doubling of both periods (two coupling constants). The $BC_1$ elliptic model degenerates to $A_1$ elliptic model characterized by the Lam\'e Hamiltonian. It is shown that in the space of $BC_1$ elliptic invariant, the potential becomes a rational function, while the flat space metric becomes a polynomial. The model possesses the hidden $sl(2)$ algebra for arbitrary coupling constants: it is equivalent to $sl(2)$-quantum top in three different magnetic fields. It is shown that there exist three one-parametric families of coupling constants for which a finite number of polynomial eigenfunctions (up to a factor) occur.
\end{abstract}


\vskip 2cm


\pacs{}

\maketitle

The Calogero-Moser-Sutherland models represent a remarkable family of Weyl-invariant integrable
systems (with rational, trigonometric/hyperbolic, elliptic potentials), both classical and quantum (see for review and discussions \cite{Olshanetsky:1983}).
These models appear in different physical sciences, in particular, in theory of random matrices
(see e.g. \cite{BLA:1993}) and in quantum field theory (see e.g. \cite{Nekrasov}).
In the quantum case, at least some of these models have the outstanding property of (quasi)-exact-solvability when a number of eigenstates can be found explicitly
(algebraically). Their gauge-rotated Hamiltonians, written in certain Weyl-invariant variables, are algebraic operators - specifically, differential operators with polynomial coefficients. It is worth noting that the $BC_n$ Calogero-Moser-Sutherland model is a particular case of the Inozemtsev model \cite{Inozemtsev:1989} which is seen as the most general $BC_n$ Weyl-invariant integrable system in ${\bf R}^n$.
Both $BC_n$ (elliptic) Calogero-Moser-Sutherland and $BC_n$ (elliptic) Inozemtsev quantum models were extensively studied in \cite{DGU:2001}, \cite{Brihaye:2003} and \cite{Takemura:2002} (see references therein), respectively.

Following the formal definition, any one-dimensional dynamics is integrable. Amongst Calogero-Moser-Sutherland models, there exist only two models, $A_1$ and $BC_1$,
which describe one-dimensional dynamics (in the case of $A_1$, it is the dynamics of the relative motion). A natural question to ask is what distinguishes these two models from all other integrable one-dimensional models. The goal of this paper is to show that both $A_1$ and $BC_1$ elliptic quantum systems are equivalent to $sl(2)$ quantum top in a constant magnetic field. They are quasi-exactly-solvable. The spectra of $BC_1$ elliptic model is also studied.

The $BC_{1}$ quantum elliptic model, as it was introduced in Olshanetsky-Perelomov \cite{Olshanetsky:1976}, is described by the Hamiltonian
\begin{equation}
\label{OPHam}
 {\cal H}^{(e)}_{BC_1}\ =\ -\ \half \frac{\pa^{2}}{\pa x^{2}}\ +\
 \ka_2\ \wp (2x)\ +\ \ka_3\ \wp (x)\ \equiv\ - \half \De^{(1)} + V\ ,
\end{equation}
where $\De^{(1)}$ is one-dimensional Laplace operator, $\ka_{2,3}$ are coupling constants. The Weierstrass function $\wp (x) \equiv \wp (x|g_2,g_3)$ (see e.g. \cite{WW:1927}) is defined as
\begin{equation}
\label{wp}
    (\wp'(x))^2\ =\ 4\ \wp^3 (x) - g_2\ \wp (x)\ -\ g_3\ =\ 4(\wp (x) -e_1)(\wp (x) -e_2)(\wp (x) -e_3),
\end{equation}
where $g_{2,3}$ are its invariants and $e_{1,2,3}$ are roots, $e_1+e_2+e_3=0$. If one of the coupling constants vanishes, $\ka_{2}(\ka_{3})=0$, the Hamiltonian becomes $(A_1)$-Lam\'e Hamiltonian (see e.g. \cite{Lame:1989} and references therein). If in the elliptic potential of Eq. (\ref{OPHam}) the trigonometric limit is taken (one of periods tends to infinity which implies the condition  $\De \equiv g_2^3 + 27 g_3^3 = 0$ holds) the Hamiltonian of $BC_1$ trigonometric/hyperbolic or generalized P\"oschl-Teller model emerges.

Since we will be interested in the general properties of the operator ${\cal H}^{(e)}_{BC_1}$, without a loss of generality, we assume that the operator (\ref{OPHam}) is defined on real line, $x \in {\bf R}$ and for the sake of convenience the fundamental domain of the Weierstrass function is rectangular with real period 1 and imaginary period $i\ \tau$. The discrete symmetry of the Hamiltonian (\ref{OPHam}) is $\mathbb{Z}_2 \oplus T_r \oplus T_c$. It consists of reflection $\mathbb{Z}_2 (x \rar -x)$ which is $BC_1$ Weyl group and two translations $T_r:\ x \rar x + 1$ and $T_c:\ x \rar x + i\ \tau$ (periodicity). Perhaps, $\mathbb{Z}_2 \oplus T_r \oplus T_c$ can make sense as double-affine $BC_1$ Weyl group.

We will consider a formal eigenvalue problem
\begin{equation}
\label{sp}
     {\cal H}^{(e)}_{BC_1} \Psi\ =\ E \Psi\ ,
\end{equation}
without posing concrete boundary conditions. It can be immediately checked that (\ref{sp}) has the exact solution
\begin{equation}
\label{gs}
 \Psi_{0}\ =\ [\ \wp'(x)\ ]^{\mu}\ ,
\end{equation}
for coupling constants
\begin{equation}
\label{par0}
      \ka_2\ =\ 2\mu(\mu-1)\ ,\ \ka_3\ =\ 2\mu(1+2\mu)\ ,
\end{equation}
and $\mu$ is an arbitrary parameter, for which the eigenvalue
\[
   E_0\ =\ 0\ .
\]
It implies that for parameters (\ref{par0}) the Hamiltonian ${\cal H}^{(e)}_{BC_1}$ has one-dimensional invariant subspace, $\Psi_{0}$ has the meaning of zero mode, and if $x \in [0,1]$, the function $\Psi_{0}$ (\ref{gs}) is the ground state function (no nodes).

Now let us introduce a new variable,
\begin{equation}
\label{tau}
 \ta \ =\ \wp(x)\ ,
\end{equation}
cf. \cite{Lame:1989} and references therein.
It is evident that $\ta$ is {\it invariant} with respect to the action of the group $\mathbb{Z}_2 \oplus T_r \oplus T_c$  - double affine $BC_1$ Weyl group. The first observation is that the potential (\ref{OPHam}) being written in $\ta$-variable is a rational function,
\[
 V(\ta)\ =\  \frac{\ka_2 + 4 \ka_3}{4}\ \ta\ +\
 \frac{\ka_2}{16}\ \frac{12 g_2 \ta^2 +36 g_3 \ta + g_2^2}{4 \ta^3 - g_2 \ta\ -\ g_3}
\]
and the ground state function (\ref{gs}) becomes
\begin{equation}
\label{gstau}
 \Psi_{0}(\ta)\ =\ (4 \ta^3 - g_2 \ta\ -\ g_3)^{\frac{\mu}{2}}\ ,
\end{equation}
(cf. (\ref{OPHam})), which is the determinant of the metric with upper indices (see below) to the power $\frac{\mu}{2}$. Making the gauge rotation
\[
h^{(e)}\ =\ -2(\Psi_{0})^{-1}\,{\cal H}^{(e)}_{\rm BC_1}\,\Psi_{0}
\]
and changing variable to $\ta$, we arrive at the algebraic operator
\begin{equation}
\label{htauD}
    h^{(e)}(\ta)\ =\ \De_g(\ta)\ +\
    {\mu} (12\ta^2-g_2) \pa_{\ta}\ -\ {\tilde\ka_3}\ta
\end{equation}
where $\De_g$ is one-dimensional Laplace-Beltrami operator
\[
    \De_g(\ta)\ =\ g^{-1/2} \ \frac{\pa}{\pa \ta} g^{1/2} g^{11} \frac{\pa}{\pa \ta}\ =\
   g^{11} \frac{\pa^2}{\pa \ta^2} + \frac{g^{11}_{,1}}{2} \frac{\pa}{\pa \ta}
\]
with flat metric
\[
g^{1 1}\ =\ \ (4 \ta^3 - g_2 \ta\ -\ g_3) \ =\ \frac{1}{g}\ ,
\]
here $g$ is its determinant with upper indices, and
\[
    \tilde\ka_3 \ =\ 2\ka_3 - 4 \mu(1+2\mu)\ \equiv \ 2n(2n+1+6\mu)\ .
\]
In the explicit form the gauge-rotated operator (\ref{htauD}) looks like
\begin{equation}
\label{htauA}
    h^{(e)}(\ta)\ =\ \ (4 \ta^3 - g_2 \ta\ -\ g_3) \pa^2_{\ta} \ +\
    (1+2\mu) (6\ta^2-\frac{g_2}{2}) \pa_{\ta}\ -\ 2n(2n+1+6\mu) \ta\ .
\end{equation}
It can be easily checked that if parameter $n$ is a non-negative integer, the operator $h^{(e)}(\ta)$ (\ref{htauA}) has the invariant subspace
\[
     {\cal P}_{n}\ =\ \langle \ta^{p} \vert \ 0 \le p \le n \rangle\ ,
\]
of dimension
\[
      \dim {\cal P}_{n}\ =\ (n+1)\ ,
\]
namely,
\[
  h^{(e)}\ :\ {\mathcal P}_{n} \ \mapsto \ {\mathcal P}_{n}\ .
\]

The space ${\cal P}_{n}$ is invariant w.r.t. $1D$ projective (M\"obius) transformation
\[
   \ta\ \mapsto \ \frac{a\ \ta + b}{c\ \ta + d}\ .
\]
Furthermore, the space ${\cal P}_{n}$ is the finite-dimensional representation space of the algebra $sl(2)$ of the first order differential operators realized as
\begin{equation}
\label{sl2}
    {\cal J}^+(n)\ =\ \ta^2\pa_{\ta} - n \ta\ ,\ {\cal J}^0(n)\ =\ \ta\pa_{\ta} - n\ ,\ {\cal J}^-(n)\ =\ \pa_{\ta}\ .
\end{equation}
Hence, the operator (\ref{htauD}) can be rewritten in terms of $sl(2)$-generators
\[
 h^{(e)}\ =\ 4\ {\cal J}^+(n)\ {\cal J}^0(n)\ -\ g_2 {\cal J}^0(n)\ {\cal J}^- - \ g_3\ {\cal J}^-\  {\cal J}^-
\]
\begin{equation}
\label{htau-sl2}
 +\ 2\ \big(4n + 1 + 6\mu \big)\ {\cal J}^+(n)\ -\ {g_2}\ \big(n+\frac{1}{2}+\mu \big){\cal J}^- \ .
\end{equation}
Thus, it is $sl(2)$ quantum top in a constant magnetic field.
This representation holds for any value of $n$. Thus, the algebra $sl(2)$ is the hidden algebra of $BC_1$ elliptic model with arbitrary coupling constants $\ka_{2,3}$ parametrized as follows
\begin{equation}
\label{par}
      \ka_2\ =\ 2\mu(\mu-1)\ ,\ \ka_3\ =\ (n + 2\mu)\ (n + 2\mu + 1)\ .
\end{equation}
If $n$ takes an integer value, the hidden algebra $sl(2)$ appears in finite-dimensional representation, and the operator (\ref{htauA}) has finite-dimensional invariant subspace and possesses a number of polynomial eigenfunctions $P_{n,i}(\ta; \mu)\ , \ i=1,\ldots (n+1)$. These polynomials can be called $BC_1$ Lam\'e polynomials (of the first kind). If $\mu = 0, 1$ these polynomials degenerate to Lam\'e polynomials of the first (fourth) kind, respectively. For example, for $n=0$ at
coupling constants (\ref{par0}) (or (\ref{par}) at $n=0$),
\[
    E_{0,1} = 0\ ,\ P_{0,1} = 1\ .
\]
For $n=1$ at coupling constants
\[
    \ka_2\ =\ 2\mu\ (\mu-1)\ ,\ \ka_3\ =\ 2(1+2\mu)\ (1+\mu)\ ,
\]
the eigenstates are
\[
   E_{\mp} = \pm (1+2\mu) \sqrt {3 g_2}\ ,\ P_{1,\mp}\ =\ \ta \mp \frac{1}{2} \sqrt{\frac{g_2}{3}}
\]
As a function of $g_2$ both eigenvalues (eigenfunctions) are branches of double-sheeted Riemann surface. Note that if $\mu=-\frac{1}{2}$ degeneracy occurs: both eigenvalues coincide, they are equal to zero, any linear function is an eigenfunction. If $g_2=0$ but $\mu \neq -\frac{1}{2}$, the Jordan cell occurs: both eigenvalues are equal to zero but there exists a single eigenfunction, $P=\ta$. In general, for $n > 1$, polynomial eigenfunctions have a form of a polynomial in $\ta$ of degree $n$, they (as well as the eigenvalues) are branches of $(n+1)$-sheeted Riemann surfaces in the parameter $g_2$. To summarize, it can be stated that for coupling constants (\ref{par}) at integer $n$, the Hamiltonian (\ref{OPHam}) has $(n+1)$ eigenfunctions of the form
\begin{equation}
\label{eigen-1}
 \Psi_{n,i}\ =\ P_{n,i}(\ta; \mu)\ \Psi_0\ , \quad i=1,\ldots (n+1) \ ,
\end{equation}
where $\Psi_0$ is given by (\ref{gs}).

It can be checked that the eigenvalue problem (\ref{sp}) has an exact solution other than (\ref{gs}),
\begin{equation}
\label{gs-k}
 \Psi_{0,k}\ =\ [\ \wp'(x)\ ]^{\mu}\ \big(\wp(x) - e_k\big)^{\frac{1}{2}-\mu}\ ,
\end{equation}
for coupling constants
\begin{equation}
\label{par0-k}
      \ka_2\ =\ 2\mu(\mu-1)\ ,\ \ka_3\ =\ (1+2\mu)(1-\mu)\ ,
\end{equation}
where $\mu$ is an arbitrary parameter, for which the eigenvalue is
\[
   E_{0,k}\ =\ \frac{(4\mu^2-1)}{2} e_k\ ,
\]
here $e_k$ is the $k$th root of the Weierstrass function (\ref{wp}).
It implies that for parameters (\ref{par0-k}) the Hamiltonian ${\cal H}^{(e)}_{BC_1}$ has one-dimensional invariant subspace.

Making a gauge rotation of the Hamiltonian (\ref{OPHam}) with subtracted $E_{0,k}$,
\[
h^{(e)}_k\ =\ -2(\Psi_{0,k})^{-1}\,({\cal H}^{(e)}_{\rm BC_1}-E_{0,k}) \,\Psi_{0,k}
\]
and changing variable to $\ta$, we arrive at the algebraic operator
\begin{equation}
\label{htauA-k}
    h^{(e)}_k(\ta)\ =\ (\ta - e_k)^{-\half + \mu}\ (h^{(e)}(\ta)-2E_{0,k})\ (\ta - e_k)^{\half - \mu}
    \ =\ \
\end{equation}
\[
(4 \ta^3 - g_2 \ta\ -\ g_3) \pa^2_{\ta}
\ +\
    2\big((5+2\mu) \ta^2 + 2(1-2\mu) e_k (\ta + e_k) - (3-2\mu)\frac{g_2}{4}\big) \pa_{\ta}
\]
\[
\ -\ 2 \tilde \ka_3 \ta\ ,
\]
(cf. (\ref{htauA})), where $e_k$ is $k$th root of the Weierstrass function (see (\ref{wp})), and
$$\tilde \ka_3 \ = \ \ka_3 - (1 - \mu)(1+2\mu)\ .$$
It can be checked that if $\tilde \ka_3 \ = \ 2n(n-1)+n(2\mu+5)$ and the parameter $n$ takes non-negative integer values, the operator $h^{(e)}_k(\ta)$ has the invariant subspace ${\cal P}_{n}$. Furthermore, the operator (\ref{htauA-k}) can be rewritten in terms of $sl(2)$-generators (\ref{sl2}) for any value of $n$, cf. (\ref{htau-sl2}),
\[
 h^{(e)}_k\ =\ 4\ {\cal J}^+(n)\ {\cal J}^0(n)\ -\ g_2\ {\cal J}^0(n)\ {\cal J}^-\
 - \ g_3\ {\cal J}^-\ {\cal J}^-
\]
\begin{equation}
\label{htau-sl2-k}
 + 2(4n + 3 + 2\mu) {\cal J}^+(n)\
 + \ 4(1-2\mu)e_k ({\cal J}^0(n) + n)
 + \ 2 \big(2(1-2\mu)e_k^2 - (2n + 3 - 2\mu)\frac{g_2}{4}\big){\cal J}^- \ .
\end{equation}
Thus, it is $sl(2)$ quantum top in constant magnetic field.

Hence, the algebra $sl(2)$ is the hidden algebra of $BC_1$ elliptic model with arbitrary coupling constants $\ka_{2,3}$ parametrized as follows
\begin{equation}
\label{par-k}
      \ka_2\ =\ \mu(\mu-1)\ ,\ \ka_3\ =\ 2n^2 + n(3+2\mu) + (1+2\mu)(1-\mu)\ ,
\end{equation}
(cf. (\ref{par})).
If $n$ takes an integer value, the hidden algebra $sl(2)$ appears in a finite-dimensional representation,
the operator (\ref{htauA-k}) has a finite-dimensional invariant subspace and possesses a number of polynomial eigenfunctions $P_{n,i}(\ta; \mu, e_k )\ , \ i=1,\ldots (n+1)$ and $k=1,2,3$. These polynomials can be called $BC_1$ Lam\'e polynomials (of the second kind). If $\mu = 0, 1$ these polynomials degenerate to Lam\'e polynomials of the second (third) kind, respectively. For example, for $n=0$ at couplings (\ref{par-k}),
\[
    E_{0,1}\ =\ \frac{(4\mu^2-1)}{2} e_k\ ,\ P_{0,1} = 1\ .
\]

In general, for $n > 1$, polynomial eigenfunctions have a form of a polynomial in $\ta$ of degree $n$, they (as well as the eigenvalues) are branches of $(n+1)$-sheeted Riemann surfaces in $g_2$.
To summarize, it can be stated that for coupling constants (\ref{par-k}), and at integer $n$, the Hamiltonian
(\ref{OPHam}) has $(n+1)$ eigenfunctions of the form
\begin{equation}
\label{eigen-2}
 \Psi_{n,i;k}\ =\ P_{n,i}(\ta; \mu, e_k)\ \Psi_{0,k}\ , \quad i=1,\ldots (n+1) \ ,\ k=1,2,3\ ,
\end{equation}
where $\Psi_{0,k}$ is given by (\ref{gs-k}).


It can be checked that the eigenvalue problem (\ref{sp}) has one more exact solution other than (\ref{gs}) or (\ref{gs-k}),
\begin{equation}
\label{gs-ij}
 \Psi_{0,\tilde k}\ =\ [\ \wp'(x)\ ]^{\nu}\  [(\wp(x) - e_i) (\wp(x) - e_j)]^{\frac{1}{2}-\nu}\ ,
\end{equation}
where $\tilde k$ is complement to $(i,j)$, for coupling constants
\begin{equation}
\label{par0-ij}
      \ka_2\ =\ 2\nu(\nu-1)\ ,\ \ka_3\ =\ \nu(1-\nu)\ ,
\end{equation}
where $\nu$ is an arbitrary parameter, for which the eigenvalue is
\[
   E_{0,\tilde k}\ =\ \frac{(1-2\nu)(3-2\nu)}{2} e_k\ ,
\]
here $e_k$ is the $\tilde k$th root of the Weierstrass function (\ref{wp}).
It implies that for parameters (\ref{par0-ij}) the Hamiltonian ${\cal H}^{(e)}_{BC_1}$ has one-dimensional invariant subspace. If in (\ref{gs-ij}) $\nu=1-\mu$ the solution (\ref{gs-k}) occurs.

Making a gauge rotation of the Hamiltonian (\ref{OPHam}) with subtracted $E_{0,\tilde k}$,
\[
h^{(e)}_{\tilde k}\ =\ -2(\Psi_{0,\tilde k})^{-1}\,({\cal H}^{(e)}_{\rm BC_1}-E_{0,\tilde k}) \,\Psi_{0,\tilde k}
\]
and changing variable to $\ta$, we arrive at the algebraic operator
\begin{equation}
\label{htauA-ij}
    h^{(e)}_{\tilde k}(\ta)\ =\ [(\ta - e_i)(\ta - e_j)]^{-\half + \mu}\ (h^{(e)}(\ta)-2E_{0,\tilde k})\ [(\ta - e_i)(\ta - e_j)]^{\half - \mu}
    \ =\ \
\end{equation}
\[
(4 \ta^3 - g_2 \ta\ -\ g_3) \pa^2_{\ta}
\ +\
    2\big((7-2\nu) \ta^2 + 2(2\nu-1) e_k (\ta + e_k) - (5+2\nu)\frac{g_2}{4}\big) \pa_{\ta}
\]
\[
\ -\ 2 \tilde \ka_3 \ta\ ,
\]
(cf. (\ref{htauA})), where $e_k$ is $k$th root of the Weierstrass function, see (\ref{wp}) and
$$\tilde \ka_3 \ = \ \ka_3 - \nu (3-2\nu)\ .$$
It can be checked that if $\tilde \ka_3 \ = \ 2n(n-1)+n(7-2\nu)$, and the parameter $n$ takes a non-negative integer value, the operator $h^{(e)}_{\tilde k}(\ta)$ has the invariant subspace ${\cal P}_{n}$. Furthermore, the operator (\ref{htauA-ij}) can be rewritten in terms of $sl(2)$-generators (\ref{sl2}) for any value of $n$, cf. (\ref{htau-sl2}),
\[
 h^{(e)}_{\tilde k}\ =\ 4\ {\cal J}^+(n)\ {\cal J}^0(n)
 -\ g_2 {\cal J}^0(n)\ {\cal J}^- - \ g_3\ {\cal J}^-\ {\cal J}^-
\]
\begin{equation}
\label{htau-sl2-tildek}
 + \  2(4n + 5 - 2\nu)\ {\cal J}^+(n)\
 + \ 4(2\nu-1)e_k ({\cal J}^0(n) + n)
 + \ 2 \big(2(2\nu-1)e_k^2 - (2n + 1 + 2\nu)\frac{g_2}{4}\big){\cal J}^- \ .
\end{equation}
Thus, it is $sl(2)$ quantum top in constant magnetic field.

Hence, the algebra $sl(2)$ is the hidden algebra of $BC_1$ elliptic model with arbitrary coupling constants $\ka_{2,3}$ parametrized as follows
\begin{equation}
\label{par-ij}
      \ka_2\ =\ \nu(\nu-1)\ ,\ \ka_3\ =\ 2n^2 + n(5-2\nu) + \nu(1-2\nu)\ ,
\end{equation}
(cf. (\ref{par}), (\ref{par-k})).
If $n$ takes an integer value, the hidden algebra $sl(2)$ appears in finite-dimensional representation, and the operator (\ref{htauA-ij}) has finite-dimensional invariant subspace
${\cal P}_{n}$ and possesses a number of polynomial eigenfunctions $\tilde P_{n,i}(\ta; \nu, e_k )\ , \ i=1,\ldots (n+1)$ and $k=1,2,3$. These polynomials can be called $BC_1$ Lam\'e polynomials (of the third kind). If $\nu = 0, 1$ these polynomials degenerate to Lam\'e polynomials of the third (second) kind, respectively. For example, for $n=0$ at couplings (\ref{par-ij}),
\[
    E_{0,1}\ =\ \frac{(1-2\nu)(3-2\nu)}{2} e_k\ ,\ P_{0,1} = 1\ .
\]

In general, for $n > 1$, the polynomial eigenfunctions have a form of a polynomial in $\ta$ of degree $n$, and they (as well as the eigenvalues) are branches of $(n+1)$-sheeted Riemann surface in $g_2$.
To summarize, it can be stated that for coupling constants (\ref{par-ij}) at integer $n$ the Hamiltonian (\ref{OPHam}) has $(n+1)$ eigenfunctions of the form
\begin{equation}
\label{eigen-3}
 \tilde \Psi_{n,i;k}\ =\ \tilde P_{n,i}(\ta; \nu, e_k)\ \Psi_{0,k}\ , \quad i=1,\ldots (n+1) \ ,\ k=1,2,3\ ,
\end{equation}
where $\Psi_{0,\tilde k}$ is given by (\ref{gs-ij}).

\noindent
{\it Observation:} Let us construct the operator
\[
       i_{par}^{(n)}(\ta)\ =\ \prod_{j=0}^n ({\cal J}^0(n) + j)\ ,
\]
where ${\cal J}^0(n)$ is the Euler-Cartan generator of the algebra $sl(2)$ (\ref{sl2}). It can be shown that any algebraic operator $h^{(e)}$ (\ref{htau-sl2}), (\ref{htau-sl2-k}), (\ref{htau-sl2-tildek}) at integer $n$ commutes with $i_{par}^{(n)}(\ta)$,
\[
  [h^{(e)}(\ta)\ ,\ i_{par}^{(n)}(\ta)]:\ {\cal P}_{n} \ \mapsto \ 0\ ,
\]
Hence, $i_{par}^{(n)}(\ta)$ is the particular integral \cite{Turbiner:2013p} of the $BC_1$ elliptic model (\ref{OPHam}).

In this paper we demonstrate that $BC_1$ elliptic model belongs to one-dimensional quasi-exactly-solvable (QES) problems \cite{Turbiner:1988}. However, it is not in
the list of known QES problems (see e.g. \cite{Turbiner:1994}). We show the existence of three different algebraic forms of the $BC_1$ Hamiltonian - all of them are the second order polynomial elements of the universal enveloping algebra $U_{sl(2)}$. If this algebra appears in a finite-dimensional representation those elements possess a finite-dimensional invariant subspace. This phenomenon occurs for any of three one-parametric subfamilies of coupling constants for which polynomial eigenfunctions may occur.
It is worth noting that a certain algebraic forms for a general $BC_n$ elliptic model were found some time ago in \cite{DGU:2001, Brihaye:2003}.

\begin{acknowledgments}
  This work was supported in part by the University Program FENOMEC, and by the PAPIIT
  grant {\bf IN109512} and CONACyT grant {\bf 166189}~(Mexico).
\end{acknowledgments}

\end{document}